\documentclass[twocolumn,showpacs,prd]{revtex4}
\usepackage{mathrsfs,bm,multirow}
\usepackage{longtable,lscape}
\usepackage{txfonts}
\usepackage{amssymb}
\usepackage{indentfirst}
\usepackage{graphicx,,booktabs}
\usepackage{color}
\usepackage{amssymb}

\usepackage[dvipdfm,  %pdftex,pdflatex
            ]{hyperref}

\begin{document}

\title{$\eta$ transitions between charmonia with meson loop contributions}
\author{Dian-Yong Chen$^{1,3}$}\email{chendy@impcas.ac.cn}
\author{Xiang Liu$^{1,2}$\footnote{Corresponding author}}\email{xiangliu@lzu.edu.cn}
\author{Takayuki Matsuki$^4$}\email{matsuki@tokyo-kasei.ac.jp}
\affiliation{$^1$Research Center for Hadron and CSR Physics, Lanzhou
University $\&$ Institute of Modern Physics of CAS, Lanzhou 730000,
China
\\$^2$School of Physical Science and Technology, Lanzhou University, Lanzhou 730000, China
\\$^3$Institute of Modern Physics, Chinese Academy of Sciences, Lanzhou 730000, China
\\$^4$Tokyo Kasei University, 1-18-1 Kaga, Itabashi, Tokyo 173-8602, Japan}

\begin{abstract}
We study the $\eta$ transitions between $\psi(4040/4160)$
and $J/\psi$ by introducing charmed meson loops in an effective
Lagrangian approach to enhance the decay amplitudes.
The branching fractions $\mathcal{B}[\psi(4040) \to J/\psi \eta]$ and
$\mathcal{B}[\psi(4160) \to J/\psi \eta]$
estimated in this paper can remarkably explain the experimental measurements of Belle and
BESIII within a reasonable parameter range. The $\eta^\prime$
transition between $\psi(4160)$ and $J/\psi$ is also investigated,
and the branching fraction is under the upper limit of CLEO, which
can be tested by future experiments.

\end{abstract}

\maketitle

\section{Introduction}
Among many heavy quarknonia, the chamonium sector especially has much abundant
spectroscopy and decay modes observed \cite{Beringer:1900zz}.
In their mass range, there are also many discrepancies
between theoretical predictions and experimental measurements
because of the non-pertubative property of Quantum Chromodynamics (QCD).
The study of spectroscopy and decay behavior of the charmonia
undoubtedly enrich our knowledge of how QCD works in hadron physics.

When considering the decay process $\psi_{1}\to \psi_{2}
\mathcal{P}$, the direct coupling among the initial and final
charmonia, $\psi_{1}$ and $\psi_{2}$, and a light chiral meson,
$\mathcal{P}$, is highly suppressed due to the OZI rule.
Because Belle and BESIII have recently observed branching fractions
for some processes of this kind, of the order of $10^{-3}$, hence we need to
consider other mechanism which enhances this type of decay amplitude. The
charmonia, whose masses are above the threshold of a charmed meson
pair, dominantly decay into a pair of charmed mesons, which can
couple with a light meson and a charmonium by exchanging a proper
charmed meson in the final states. The contribution from such a
structure, i.e., meson loop contribution, is dominant in this decay
and becomes important in understanding decay behaviors of higher
charmonia.

Taking $\psi(3770)$ as an example, which is the first charmonium
above the threshold of open charmed mesons, the BES Collaboration
announced that the branching fraction of its non-$D\bar{D}$ decay is
$\mathcal{B}[\psi(3770) \to \mathrm{non}-D\bar{D}]=(14.7\pm 3.2) \%$
\cite{Ablikim:2006zq, Ablikim:2006aj, Ablikim:2008zzb,
Ablikim:2007zz}. Such a large non-$D\bar{D}$ branching fraction is
several times larger than expected in theory \cite{He:2008xb}. To
resolve this discrepancy,
% between the
%experimental measurements and theoretical predictions,
the authors in Refs. \cite{Liu:2009dr,Zhang:2009kr} have taken
account of the charmed meson loop, where the initial
charmonium $\psi(3770)$ decays into a charmed meson pair $D\bar{D}$
and this pair couples to a vector and/or pseudoscalar
meson by exchanging a $D$ meson. After including the meson loop
contributions, the large non-$D\bar{D}$ branching fraction of
$\psi(3770)$ can be nicely explained.
%{\color{red}
The decay mode
$\psi(3770) \to J/\psi \mathcal{P}$ and the lineshape around
$\psi(3770)$ have been studied with meson loop contributions in Ref.
\cite{Wang:2011yh}.
%%%%%%%%%
\iffalse
Because $\eta/\eta'$ transition is nothing but the hidden charm decay,
the decay structure is very similar to the one we have studied on
decays of $X(3915)$ and $Z(3940)$ in Ref. \cite{chen2012}, where we have
assumed assignment, two charmonium-like states $X(3915)$ and $Z(3930)$ for
$\chi_{c0}^\prime(2P)$ and $\chi_{c2}^\prime (2P)$
and have succeeded in explaining the fact that the only one structure
$X(3915)$ has been observed in the $J/\psi\omega$ invariant mass spectrum
for the process $\gamma\gamma\to J/\psi\omega$.
\fi
%%%%%%%%%

Even though the measurements of the higher charmonia are not yet
enough to discuss $\eta$ transitions, there appear some
experiments of this kind. In the International Conference on High
Energy Physics, the Belle Collaboration reported their
measurements for $\eta$ transitions between $\psi(4160/4040)$ and
$J/\psi$ \cite{Bruce:2012,Wang:2012bg}. They announced that
$\mathcal{B}[\psi(4040) \to \eta J/\psi] \cdot
\Gamma_{e^+e^-}(\psi(4040))= 4.8 \pm 0.9 \pm 1.4 $ eV or $11.2 \pm
1.3 \pm 1.9$ eV with different fitting parameters to the data. The
corresponding results for $\psi(4160)$ are $\mathcal{B}[\psi(4160)
\to \eta J/\psi] \cdot \Gamma_{e^+e^-}(\psi(4160))= 4.0 \pm 0.8
\pm 1.4$ eV or $13.8 \pm 1.3 \pm 2.0$ eV. Taking $\Gamma_{e^+ e^-}
(\psi(4040)) =(0.86 \pm 0.07)$ keV and $ \Gamma_{e^+
e^-}(\psi(4160)) =(0.83 \pm 0.07)$ keV, one obtains the branching
ratios as $\mathcal{B}[\psi(4040) \to J/\psi \eta]=(0.56 \pm 0.10
\pm 0.17) \% $ or $(1.30 \pm 0.15 \pm 0.24) \%$ and
$\mathcal{B}[\psi(4160) \to J/\psi \eta]= (0.48 \pm 0.10 \pm
0.17)\%$ or $(1.66 \pm 0.16 \pm 0.28)\%$. Before this measurement,
only the upper limits of branching ratios for $\psi(4040) \to \eta
J/\psi$ and $\psi(4160)\to \eta J/\psi $ were reported by the CLEO
Collaboration, which are $<7\times 10^{-3}$ and $<8 \times
10^{-3}$ \cite{Coan:2006rv}, respectively. Recently, the BESIII
Collaboration also analyzed the production of $e^+ e^- \to \eta
J/\psi$ at a center-of-mass energy of $\sqrt{s}=4.009$ GeV.
Because the Born cross section is reported to be $(32.1 \pm 2.8
\pm 1.3)$ pb \cite{Ablikim:2012ht}, the corresponding fractional
transition rate is $\mathcal{B}\left[\psi(4040) \to \eta
J/\psi\right] =(5.2 \pm 0.5 \pm 0.2 \pm 0.5) \times 10^{-3}$,
which is consistent with the first solution of the Belle
Collaboration and measurement by CLEO. With regard to experimental
data, in this paper we will use the old data by CLEO
\cite{Coan:2006rv} and others as well as the most recent data
given by Belle \cite{Wang:2012bg} and BESIII
\cite{Ablikim:2012ht}. This is because PDG has not yet included
the most recent data by Belle and BESIII.
%Before the measurement by Belle, the CLEO collaboration obtained the upper
%limits of $\psi(4040)\to J/\psi\eta$ and $\psi(4160)\to
%J/\psi\eta$ as shown in Ref. \cite{Coan:2006rv}. At present, PDG
%\cite{Beringer:1900zz} only adopted the CLEO's results for
%$\psi(4040)\to J/\psi\eta$ and $\psi(4160)\to J/\psi\eta$. The
%Belle Collaboration reported their results of $\psi(4040)\to
%J/\psi\eta$ and $\psi(4160)\to J/\psi\eta$ in October 2012 while
%PDG released their new PDG2012 in July 2012, which is the reason
%why PDG did not collect the new data from Belle
%\cite{Wang:2012bg}.}

Similar to the case of $\psi(3770)$, $\psi(4040)$ and $\psi(4160)$
are above the threshold of charmed meson pairs, and dominantly decay into
these. The experimental measurements stimulate us to study
the $\eta$ transition between $\psi(4040/4160)$ and
$J/\psi$ with the meson loop mechanism, which is essential to understand
the hidden charm decay behavior like higher charmonia.

This paper is organized as follows. After introduction, a brief
review of meson loop mechanism is presented and the corresponding
amplitudes are calculated using an effective Lagrangian in Section II.
Our numerical results of the branching ratios %resulted from meson loop mechanism
are given in Section III. Section IV is devoted to
summary.

%\section{Meson Loop effects on $\eta$ transitions}
\section{$\eta$ transition of $\psi(4040)$ and $\psi(4160)$}
\begin{figure}[!h]
\centering %
\scalebox{0.7}{\includegraphics{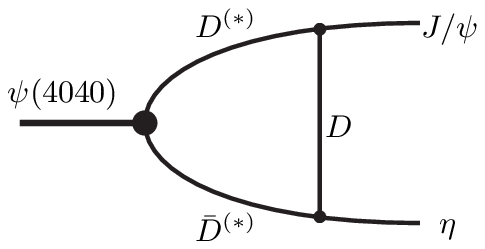}}\hspace{5mm}
\scalebox{0.7}{\includegraphics{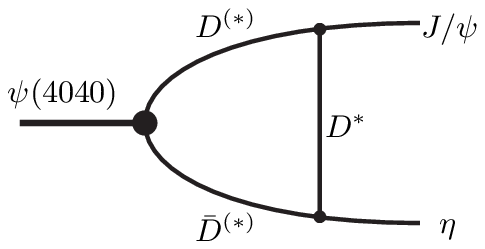}} %
\caption{The typical meson loop diagrams contributing to $\psi(4040)
\to J/\psi \eta$. The initial $\psi(4040)$ couples with a charmed
meson pair $D^{(\ast)} \bar{D}^{(\ast)}$, and by exchanging the $D$
meson (left) or $D^\ast$ meson (right), the charmed meson pair
converts itself into $J/\psi \eta$ in the final state.  \label{Fig:DLoop}}
\end{figure}

The meson loop effect plays a crucial role in understanding the
$\eta$ transition between the higher charmonium and $J/\psi$. Taking
$\psi(4040)\to J/\psi \eta$ as an example,
involvement of the meson loop in the decay
is depicted in Fig. \ref{Fig:DLoop}. The charmonium $\psi(4040)$
is decomposed into $D^{(\ast)} \bar{D}^{(\ast)}$ and by exchanging $D$
or $D^\ast$ meson, the charmed meson pair converts itself into
$J/\psi \eta$.

The effective Lagrangian approach is adopted to evaluate the meson
loop contributions to higher charmonia decay into $J/\psi \eta$ as
shown in Fig. \ref{Fig:DLoop}. Utilizing the heavy quark limit
and chiral symmetry, the effective Lagrangians, {which involves
interactions among $J/\psi$, pseudoscalar meson, and charmed mesons}, read as
\cite{Casalbuoni:1996pg,Oh:2000qr}:

\begin{eqnarray}
\mathcal{L}_{{J/\psi \mathcal{D}^{(\ast)}\mathcal{D}^{(\ast)}}}&=& i
g_{{J/\psi \mathcal{D}\mathcal{D}}} \psi_\mu \left(
\partial^\mu \mathcal{D} {\mathcal{D}}^{\dagger} - \mathcal{D}
\partial^\mu {\mathcal{D}}^{\dagger}
\right) \nonumber\\&&-g_{{J/\psi \mathcal{D}^* \mathcal{D}}}^{}
\varepsilon^{\mu\nu\alpha\beta}
\partial_\mu \psi_\nu \left(
\partial_\alpha \mathcal{D}^*_\beta {\mathcal{D}}^{\dagger}
+ \mathcal{D} \partial_\alpha {\mathcal{D}}^{*\dagger}_\beta \right)
\nonumber\\
&& -i g_{{J/\psi \mathcal{D}^\ast \mathcal{D}^\ast}} \Big\{ \psi^\mu
\big(
\partial_\mu \mathcal{D}^{*\nu} {\mathcal{D}}_\nu^{*\dagger} -
\mathcal{D}^{*\nu}
\partial_\mu {\mathcal{D}}_\nu^{*\dagger} \big) \nonumber\\
&& + \left( \partial_\mu \psi_\nu \mathcal{D}^{*\nu}
- \psi_\nu
\partial_\mu \mathcal{D}^{*\nu} \right) {\mathcal{D}}^{*\mu\dagger}  \mbox{} \nonumber\\
&& + \mathcal{D}^{*\mu}\big( \psi^\nu
\partial_\mu {\mathcal{D}}^{*\dagger}_{\nu} - \partial_\mu \psi_\nu {\mathcal{D}}^{*\nu\dagger}
\big) \Big\},\\
%\end{eqnarray}
%%
%% DDP
%\begin{eqnarray}
\mathcal{L}_{\mathcal{D}^{(\ast)}\mathcal{D}^{(\ast)}\mathcal{P}}
&=& -i g_{\mathcal{D}^*\mathcal{D}P} (\bar{\mathcal{D}}  \partial_\mu \mathcal{P}
\mathcal{D}^{*\mu}  - \bar{\mathcal{D}}^{*\mu}  \partial_\mu
\mathcal{P}  \mathcal{D} ) \nonumber\\
&& + \frac{1}{2}
g_{\mathcal{D}^*\mathcal{D}^*P}\epsilon_{\mu\nu\alpha\beta} \bar{\mathcal{D}}^{*\mu}
\partial^\nu \mathcal{P}\;  {\stackrel{\leftrightarrow}{\partial^\alpha}}\;
\mathcal{D}^{*\beta} ,
\end{eqnarray}
with $\mathcal{D}^{(\ast)}=\left(D^{(\ast) 0 }, D^{(\ast) +}
,D_s^{(\ast)+}\right)$. Considering $\eta$ and $\eta^\prime$ mixing, one
has $\mathcal{P}$ in the form,
\begin{eqnarray}
  \mathcal{P} &=&
 \left(
 \begin{array}{ccc}
\frac{\pi^0}{\sqrt{2}} + \alpha \eta + \beta \eta^\prime & \pi^{+} & K^{+}\\
\pi^{-} & -\frac{\pi^0}{\sqrt{2}}+ \alpha \eta + \beta \eta^\prime   &  K^{0}\\
 K^{-} & \bar{K}^{0} & \gamma \eta + \delta \eta^\prime
 \end{array}
 \right),
\end{eqnarray}
where
\begin{eqnarray}
\alpha= \frac{\cos \theta -\sqrt{2} \sin \theta}{\sqrt{6}},~ \beta =
\frac{\sin \theta + \sqrt{2} \cos \theta}{\sqrt{6}}, \nonumber\\
\gamma = \frac{-2 \cos \theta -\sqrt{2} \sin \theta}{\sqrt{6}} ,~
\delta= \frac{-2 \sin \theta + \sqrt{2} \cos \theta}{\sqrt{6}}
\end{eqnarray}
and we adopt $\theta=-19.1^\circ$ in the present work
\cite{Coffman:1988ve,Jousset:1988ni}.

{Since $J/\psi$ cannot decay into
charmed mesons due to the phase space restriction, we need to
consider the symmetric limit of heavy quark effective theory to determine relations among these
coupling constants.} In this limit,
the coupling constants between $J/\psi$ and charmed mesons satisfy
\cite{Achasov:1994vh,Deandrea:2003pv} $g_{J/\psi \mathcal{DD}}=
g_{J/\psi \mathcal{D}^\ast \mathcal{D}^\ast} m_D/m_{D^\ast}
=g_{J/\psi \mathcal{D}^\ast \mathcal{D}} \sqrt{m_D m_{D^\ast}}
=m_{J/\psi}/f_{J/\psi}$ with $m_{J/\psi}$ and $f_{J/\psi}$ being the
mass and decay constant of $J/\psi$.
{The adopted values of these
coupling constants are $g_{J/\psi \mathcal{DD}}=7.44$, $g_{J/\psi
\mathcal{D}^\ast \mathcal{D}^\ast}=8.00$ and $g_{J/\psi
\mathcal{D}^\ast \mathcal{D}}=3.84$ GeV$^{-1}$, which are determined
by the vector meson dominance
\cite{Achasov:1994vh,Deandrea:2003pv}.} On the other hand, the
couplings between pseudoscalar meson and charmed mesons can be
related to the gauge coupling constant $g$ by \cite{Cheng:2004ru}
$g_{\mathcal{P D^\ast D^\ast}} =g_{\mathcal{P D^\ast D}}/\sqrt{m_D
m_{D^\ast}} =2g/f_\pi$ with $g=0.59$ and $f_\pi=132$ MeV
{following Ref. \cite{Isola:2003fh},
which were obtained using the full width of $D^{*+}$
\cite{Ahmed:2001xc}}.

The coupling constants between higher charmonia, such as
$\psi(4040)$ and $\psi(4160)$, and charmed mesons are evaluated by
the partial decay width, assuming that these two charmonia dominantly
decay into $D\bar{D}$, $D^\ast \bar{D} + h.c.$, and $D^\ast
\bar{D}^\ast$ due to the phase space restriction.
{Here, the Lorentz structure of
interaction between $\psi(4040)/\psi(4160)$ and charmed mesons is the same as
that between $J/\psi$ and charmed mesons since
$\psi(4040)/\psi(4160)$ and $J/\psi$ are vector charmonia. However,
the relative sign of these coupling constants of
$\psi(4040)/\psi(4160)$ with charmed mesons cannot be constrained and
hence, in this work we need to consider the effects due to different
signs of these coupling constants.} Furthermore the BaBar
Collaboration has measured the ratios between these decay modes for
$\psi(4040)$ and $\psi(4160)$ \cite{Aubert:2009aq}, which are
${\mathcal{B}(\psi(4040)\to D\bar{D}) /{\mathcal{B}}(\psi(4040) \to
D^\ast \bar{D})}=0.24 \pm 0.05 \pm 0.12$,
${\mathcal{B}(\psi(4040)\to D^\ast \bar{D}^\ast )
/{\mathcal{B}}(\psi(4040) \to D^\ast \bar{D})}=0.18 \pm 0.14 \pm
0.03$, {
${\mathcal{B}(\psi(4160)\to D\bar{D}) /{\mathcal{B}}(\psi(4160) \to
D^\ast \bar{D}^\ast)}=0.02 \pm 0.03 \pm 0.02$, and
${\mathcal{B}(\psi(4160)\to D^\ast \bar{D})
/{\mathcal{B}}(\psi(4160) \to D^\ast \bar{D}^\ast)}=0.34 \pm 0.14
\pm 0.05$.
%
%Together with
Given the total decay width $\Gamma_{\psi(4040)} =80\pm 10$ MeV
and $\Gamma_{\psi(4160)}=103\pm 8$ MeV, one obtains
$|g_{\psi(4040) \mathcal{DD}}| =2.13 \pm 0.36$, $|g_{\psi(4040)
\mathcal{D}^\ast \mathcal{D}}|=1.70 \pm0.15  \ \mathrm{GeV}^{-1}$,
$|g_{\psi(4040) \mathcal{D}^\ast \mathcal{D}^\ast}|= 3.34 \pm 1.00
$, $|g_{\psi(4160) \mathcal{DD}}| =0.57 \pm 0.47$, $|g_{\psi(4160)
\mathcal{D}^\ast \mathcal{D}}|=0.77\pm 0.11\ \mathrm{GeV}^{-1}$,
and $|g_{\psi(4160) \mathcal{D}^\ast \mathcal{D}^\ast}|= 2.23 \pm
0.15 $. }

With the Lagrangians listed above, we can obtain the hadronic decay
amplitudes for $\psi(4040)(p_0) \rightarrowtail [D^{(\ast)}(p_1)
\bar{D}^{(\ast)} (p_2)]D^{(\ast)}(q) \rightarrowtail J/\psi(p_3)
\eta(p_4)$,
\begin{eqnarray}
%%
%%  DD D*
\mathcal{A}_{D\bar{D}}^{D^\ast} &=& (i)^3 \int \frac{d^4q}{(2\pi)^4}
\big[ig_{\psi^\prime \mathcal{D D}} \epsilon_{\psi^\prime}^\mu (ip_{1\mu}-
ip_{2\mu})\big] \big[-g_{J/\psi \mathcal{D}^\ast \mathcal{D}}   \nonumber\\&\times&
\varepsilon_{\rho \nu \alpha \beta} (ip_3^\rho)
\epsilon_{J/\psi}^\nu
(iq^\alpha)\big] \big[-ig_{D^\ast D\eta} (ip_{4 \lambda})\big] \nonumber\\
&\times &\frac{1}{p_1^2-m_D^2} \frac{1}{p_2^2 -m_D^2}
\frac{-g^{\beta \lambda} + q^\beta q^\lambda/m_{D^\ast}^2}{ q^2
-m_{D^\ast}^2}
\mathcal{F}^2(q^2,m_{D^\ast}^2),\nonumber\\
%%
%% D D* D
\mathcal{A}_{D\bar{D}^\ast}^{D} &=& (i)^3 \int \frac{d^4q}{(2\pi)^4}
\big[-g_{\psi^\prime \mathcal{D}^\ast \mathcal{D}} \varepsilon_{\rho \mu \alpha \beta}
(-ip_0^\rho) \epsilon_{\psi^\prime}^\mu (ip_2^\alpha)
\big]\nonumber\\&\times & \big[ig_{J/\psi \mathcal{DD}} \epsilon_{J/\psi}^\nu (iq_\nu
+ip_{1 \nu})\big] \big[-ig_{D^\ast D \eta} (-ip_{4 \lambda})\big] \nonumber\\
&\times& \frac{1}{p_1^2 -m_D^2} \frac{-g^{\beta \lambda} + p_2^\beta
p_2^\lambda/m_{D^\ast}^2}{p_2^2 -m_{D^\ast}^2} \frac{1}{q^2 -m_D^2}
\mathcal{F}^2(q^2,m_D^2),\nonumber\\
%%
%% D D* D*
\mathcal{A}_{D \bar{D}^\ast}^{D^\ast} &=& (i)^3 \int
\frac{d^4q}{(2\pi)^4} \big[-g_{\psi^\prime \mathcal{D}^\ast \mathcal{D}} \varepsilon_{\rho
\mu \alpha \beta} (-ip_0^\rho) \epsilon_{\psi^\prime}^\mu
(ip_2^\alpha) \big] \nonumber\\ &\times& \big[-g_{J/\psi \mathcal{D}^\ast \mathcal{D}}
\varepsilon_{\lambda \nu \theta \phi} (ip_3^\lambda)
\epsilon_{J/\psi}^\nu (iq^\theta)\big] \big[\frac{1}{2} g_{D^\ast D^\ast
\eta} \varepsilon_{\delta \tau \eta \omega} \nonumber\\ &\times&
(ip_4^\tau) (-iq^\eta +ip_2^\eta)\big] \frac{1}{p_1^2 -m_D^2}
\frac{-g^{\beta \delta}+p_2^\beta p_2^\delta/m_{D^\ast}^2}{p_2^2
-m_{D^\ast}^2} \nonumber\\ & \times&  \frac{-g^{\phi \omega} +
q^\phi q^\omega/m_{D^\ast}^2}{q^2
-m_{D^\ast}^2} \mathcal{F}^2(q^2,m_{D^\ast}^2),\nonumber\\
%%
%% D* D D*
\mathcal{A}_{D^\ast \bar{D}}^{D^\ast} &=& (i)^3 \int
\frac{d^4q}{(2\pi)^4} \big[-g_{\psi^\prime \mathcal{D}^\ast D\mathcal{}} \varepsilon_{\tau
\mu \alpha \beta} (-ip_0^\tau) \epsilon_{\psi^\prime}^\mu
(ip_1^\alpha)\big ] \nonumber\\ &\times& \big[-ig_{J/\psi \mathcal{D}^\ast \mathcal{D}^\ast}
\epsilon_{J/\psi}^\nu (g_{\rho \lambda} (-ip_{1\nu}-iq_\nu)) +
g_{\rho \nu} (ip_{3\lambda}  +ip_{1\lambda}) \nonumber\\&+& g_{\nu
\lambda}(iq_{\rho}-ip_{3 \rho})\big] \big[-ig_{D^\ast D \eta} (ip_{4
\delta})] \frac{-g^{\beta \rho}
+ p_1^\beta p_1^\rho/m_{D^\ast}^2}{p_1^2 -m_{D^\ast}^2} \nonumber\\
&\times& \frac{1}{p_2-m_D^2} \frac{-g^{\delta \lambda}+q^\delta
q^\lambda/m_{D^\ast}^2 }{q^2 -m_{D^\ast}^2} \mathcal{F}^2(q^2,
m_{D^\ast}^2),\nonumber\\
%%
%% D* D* D
\mathcal{A}_{D^\ast \bar{D}^\ast}^{D} &=& (i)^3 \int
\frac{d^4q}{(2\pi)^4}\Big[-ig_{\psi^\prime \mathcal{D}^\ast \mathcal{D}^\ast}
\epsilon_{\psi^\prime}^\mu (g_{\rho \lambda} (ip_{1\mu} -ip_{2\mu})
 \nonumber\\ &+& g_{\mu \rho} (-ip_{0 \lambda}-ip_{1\lambda})+g_{\mu
\lambda} (ip_{2\rho}+ip_{0 \rho})) \Big]\nonumber\\
&\times& \Big[-g_{J/\psi \mathcal{D}^\ast \mathcal{D}} \varepsilon_{\delta \nu \alpha \beta}
(ip_3^\delta) \epsilon_{J/\psi}^\nu (-ip_1^\alpha)\Big] \Big[-ig_{D^\ast D
\eta} (-ip_{4 \tau})\Big] \nonumber\\
&\times& \frac{-g^{\beta \rho} +p_1^\beta
p_1^\rho/m_{D^\ast}^2}{p_1^2 -m_{D^\ast}^2}  \frac{-g^{\tau \lambda}
+ p_2^\tau p_2^\lambda/m_{D^\ast}^2}{p_2^2 -m_{D^\ast}^2}
\nonumber\\&\times&\frac{1}{q^2-m_D^2} \mathcal{F}^2(q^2, m_D^2),\nonumber\\
%%
%% D* D* D*
\mathcal{A}_{D^\ast \bar{D}^\ast}^{D^\ast} &=& (i)^3 \int
\frac{d^4q}{(2\pi)^4}[-ig_{\psi^\prime \mathcal{D}^\ast \mathcal{D}^\ast}
\epsilon_{\psi^\prime}^\mu (g_{\rho \lambda} (ip_{1\mu}
-ip_{2\mu})\nonumber\\ &+& g_{\mu \rho} (-ip_{0
\lambda}-ip_{1\lambda})+g_{\mu \lambda}
(ip_{2\rho}+ip_{0 \rho})) ]\nonumber\\
&\times& \bigg[-ig_{J/\psi \mathcal{D}^\ast \mathcal{D}^\ast} \epsilon_{J/\psi}^\nu
(g_{\alpha \beta} (-ip_1^\nu -iq^\nu)) + g_{\beta \nu} (ip_{3\alpha}
+ip_{1 \alpha})\nonumber\\ &+& g_{\alpha \nu} (iq_{\beta}-ip_{3
\beta}) \bigg]\bigg [\frac{1}{2} g_{D^\ast D^\ast \eta} \varepsilon_{\delta
\tau \eta \omega} (ip_4^\tau) (-iq^\eta+ip_2^\eta)\bigg]  \nonumber\\
&\times& \frac{-g^{\rho \beta} + p_1^\beta
p_1^\rho/m_{D^\ast}^2}{p_1^2 -m_{D^\ast}^2} \frac{-g^{\lambda
\delta}+ p_2^\lambda p_2^\delta/m_{D^\ast}^2 }{p_2^2 -m_{D^\ast}^2 }
\nonumber\\
&\times&\frac{-g^{\alpha \omega} + q^\alpha q^\omega/m_{D^\ast}^2
}{q^2 -m_{D^\ast}^2} \mathcal{F}^2(q^2, m_{D^\ast}^2),
\end{eqnarray}
{In the above expressions, $g_{D^{(\ast)}D^{(\ast)}\eta}=\alpha \,g_{\mathcal{PD}^{(\ast)}\mathcal{D}^{(\ast)}}$ and
$g_{\psi(4040)\mathcal{D}^{(\ast)}\mathcal{D}^{(\ast)}}$ is abbreviated as $g_{\psi^\prime\mathcal{D}^{(\ast)}\mathcal{D}^{(\ast)}}$.}
Here, the amplitude $\mathcal{A}_{M_1 M_2}^{M_3}$ corresponds to the
process in which the initial $\psi(4040)$ is decomposed into a pair of charmed
mesons $M_1 M_2$ by exchanging $M_3$, and then this meson pair converts itself
into $J/\psi \eta$ in the final state. The form factor
$\mathcal{F}(q^2, m_E^2)=(m_E^2-\Lambda^2)/(q^2-\Lambda^2)$ is
introduced to avoid {ultraviolet} divergence in the loop integrals
as well as to describe the structure and off-shell effects
of the exchanged mesons, {and also plays a role similar to the
Pauli-Villas renormalization scheme \cite{Itzykson:1980rh,Peskin:1995ev}. Here $m_E$ denotes the mass of the exchanged charmed meson in Fig. \ref{Fig:DLoop}}, the parameter $\Lambda$ can be
reparameterized as $\Lambda= m_E + \alpha \Lambda_{QCD}$ with
$\Lambda_{QCD}=220$MeV, and the unique parameter $\alpha$ is taken
to be of the order of unity obtained in Ref. \cite{Cheng:2004ru}.

To estimate the branching ratios of $\psi^\prime \to J/\psi \eta \
(\psi^\prime=\{\psi(4040), \psi(4160)\})$ using hadronic
loops, we summarize the amplitudes as,
\begin{eqnarray}
\mathcal{M}_{1} &=& \mathcal{A}_{D \bar{D}}^{D^\ast}=
g^{DD}_{\psi^\prime J/\psi \eta} \varepsilon_{\mu \nu \alpha \beta}
\epsilon_{\psi^\prime}^{\mu}
\epsilon_{J/\psi}^\nu p_{J/\psi}^{\alpha} p_{\eta}^\beta,\label{Eq:Lorentz0}\\
\mathcal{M}_{2} &=& \mathcal{A}_{D \bar{D}^\ast}^{D} +
\mathcal{A}_{D\bar{D}^\ast}^{D^\ast} + \mathcal{A}_{D^\ast
\bar{D}}^{D^\ast} = g^{D^\ast D}_{\psi^\prime J/\psi \eta}
\varepsilon_{\mu \nu \alpha \beta} \epsilon_{\psi^\prime}^{\mu}
\epsilon_{J/\psi}^\nu p_{J/\psi}^{\alpha} p_{\eta}^\beta,\nonumber\\
\\
\mathcal{M}_{3} &=& \mathcal{A}_{D^\ast \bar{D}^\ast}^{D} +
\mathcal{A}_{D^\ast \bar{D}^\ast}^{D^\ast} =g^{D^\ast
D^\ast}_{\psi^\prime J/\psi \eta} \varepsilon_{\mu \nu \alpha \beta}
\epsilon_{\psi^\prime}^{\mu} \epsilon_{J/\psi}^\nu
p_{J/\psi}^{\alpha} p_{\eta}^\beta, \label{Eq:Lorentz}
\end{eqnarray}
where $\mathcal{M}_1$, $\mathcal{M}_2$ and $\mathcal{M}_3$ are the
amplitudes corresponding to the channels $D\bar{D}$, $D^\ast\bar{D} +h.c.$ and $D^\ast
\bar{D}^\ast$, respectively.
%After estimating the loop integrals,
The amplitudes can be reduced to a simple Lorentz structure
$\varepsilon_{\mu \nu \alpha \beta} \epsilon_{\psi^\prime}^{\mu}
\epsilon_{J/\psi}^\nu p_{J/\psi}^{\alpha} p_{\eta}^\beta$ multiplied
with the coupling constants $g^{D^{(\ast)} D^{(\ast)}}_{\psi^\prime
J/\psi \eta}$, which can be evaluated by the loop integrals.

Having the above expressions, one can obtain the total amplitudes
expressed as %%
\begin{eqnarray}
\mathcal{M}_{\mathrm{tot}}= \mathcal{M}_1 + \mathcal{M}_2 +
\mathcal{M}_3, \label{Eq:AmpTot}
\end{eqnarray}
and the partial decay width is
\begin{eqnarray}
\Gamma =\frac{1}{3} \frac{1}{8 \pi} \frac{|\vec{p}_\eta|}{
m_{\psi^\prime}^2} |\overline{\mathcal{M}_{\mathrm{tot}}}|^2,
\label{Eq:width}
\end{eqnarray}
where the overline indicates the sum over the polarization vectors
of $\psi^\prime$ and $J/\psi$ and $|\vec{p}_{\eta}|=\lambda^{1/2}
(m_{\psi^\prime}^2,m_{J/\psi}^2,m_{\eta}^2)/(2m_{\psi^\prime})$
with K\"{a}llen function $\lambda(x,y,z)=x^2+y^2+z^2-2xy-2xz-
2yz$.

\begin{figure}[h!]
\centering %
\scalebox{1}{\includegraphics{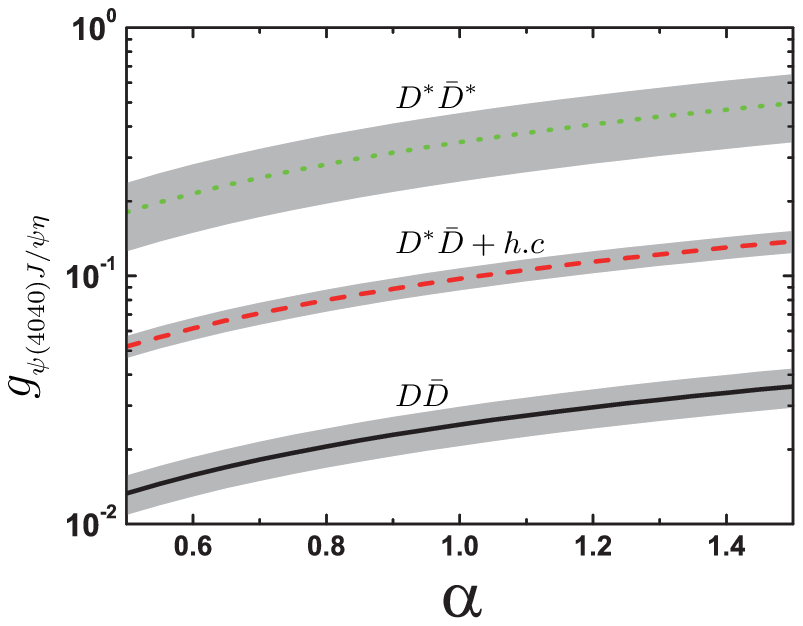}} \hspace{5mm}
\scalebox{1}{\includegraphics{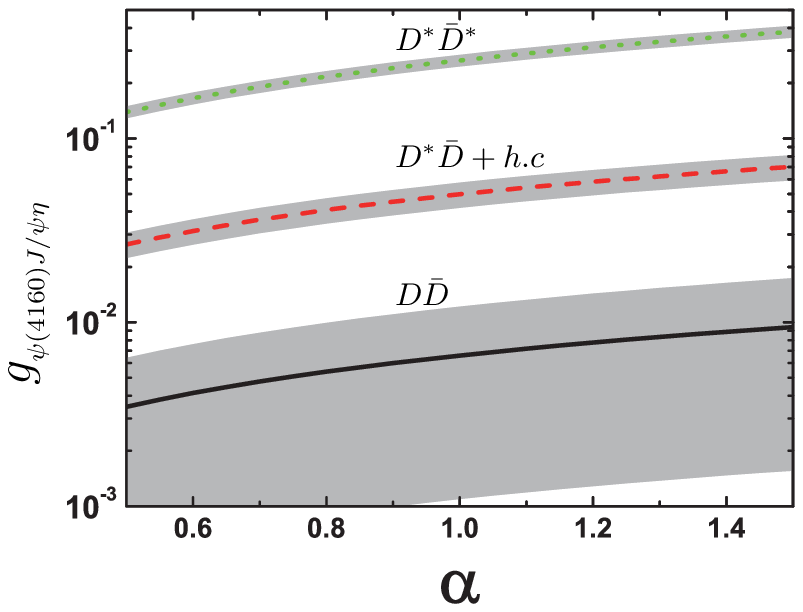}} %
\caption{{(color online) The $\alpha$ dependence of the absolute
values of coupling constants $g_{\psi^\prime J/\psi \eta} $ derived
from the meson loop contributions to $\psi(4040) \to J/\psi \eta$ (upper
panel) and $\psi(4160) \to J/\psi \eta$ (lower panel). Here, we consider different
intermediate state contributions to $g_{\psi^\prime J/\psi \eta}$, i.e., the green dotted, red dashed
and black solid curves correspond to the contributions from
intermediate $D^\ast \bar{D}^\ast$, $D^\ast \bar{D} + h.c.$ and $D\bar{D}$, respectively. The grey bands denote the uncertainties
caused by the error of the coupling between $\psi(4040)/ \psi(4160)$
and charmed meson pairs.\label{Fig:coupling}} }
\end{figure}

\begin{figure}[h!]
\centering %
\scalebox{0.725}{\includegraphics{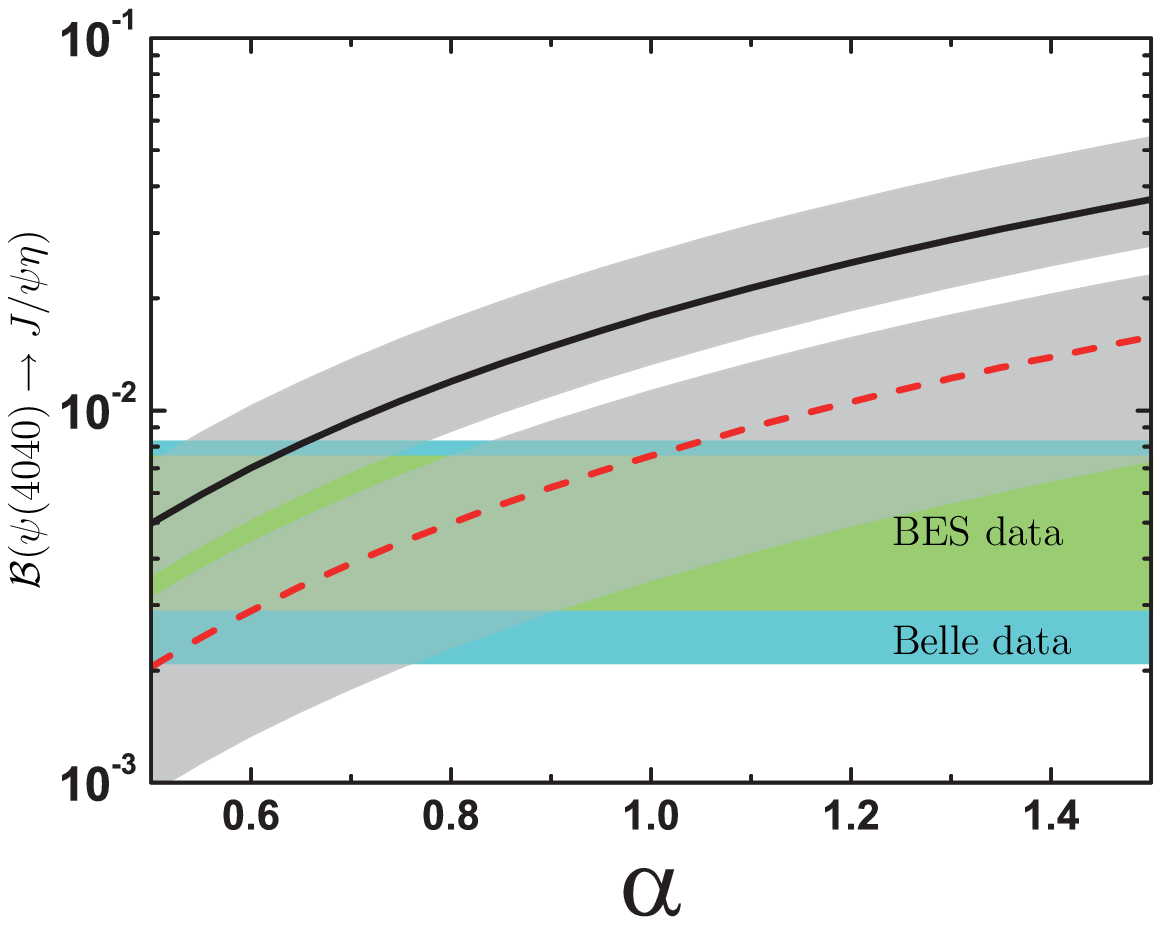}}\hspace{5mm}
\scalebox{0.7}{\includegraphics{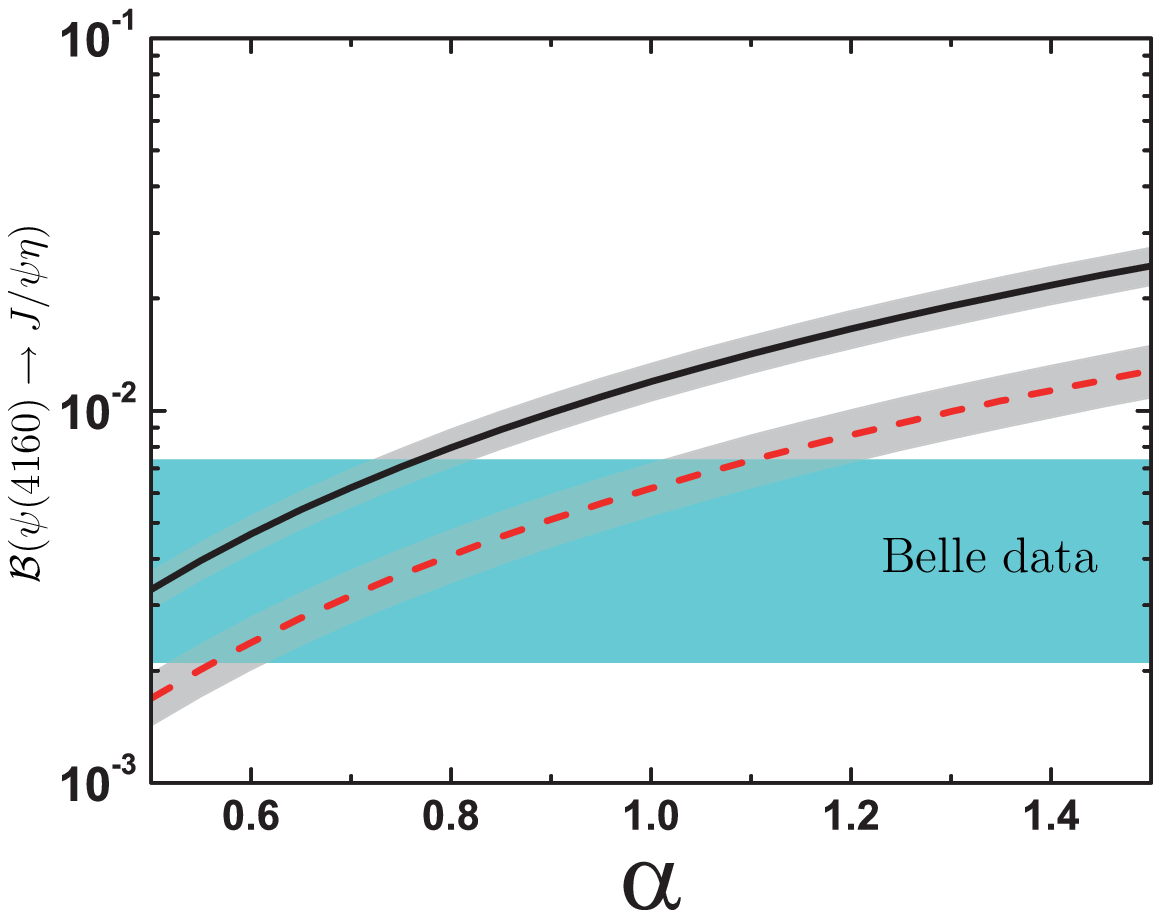}} %
\caption{{(color online) The comparison of the branching ratios
obtained for $\psi(4040) \to J/\psi \eta$ (upper panel) and
$\psi(4160) \to J/\psi \eta$ (lower panel) with the experimental
data from Belle (the cyan bands) \cite{Bruce:2012,Wang:2012bg} and
BES (the yellow band) \cite{Ablikim:2012ht}. Here, the black solid
curves with errors are the results when taking the relative sign
of $g_{\psi(4040)/\psi(4160) D^\ast D^\ast}$ and
$g_{\psi(4040)/\psi(4160) \mathcal{D}^\ast \mathcal{D}}$ as plus,
while the red dashed curves with errors correspond to the case
when the relative sign is minus. \label{Fig:branching} }}
\end{figure}

\begin{figure}[h!]
\centering %
\scalebox{1.025}{\includegraphics{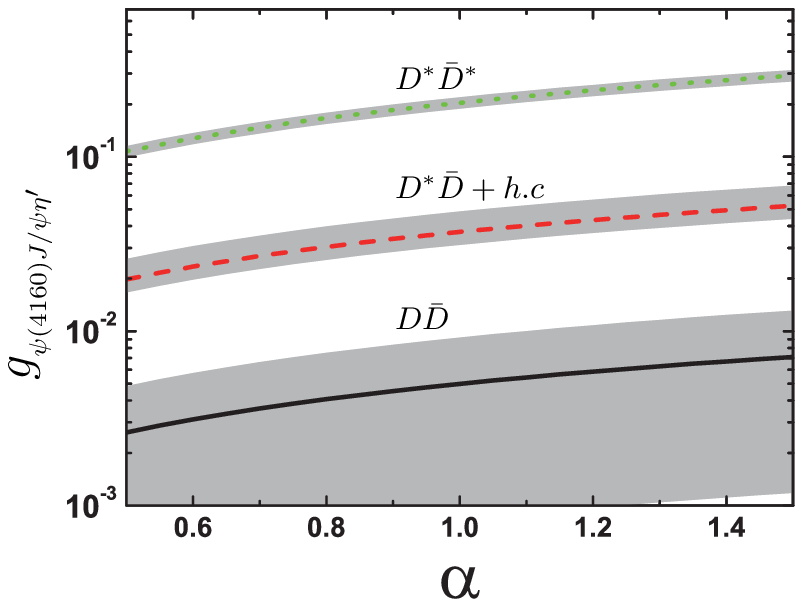}}\hspace{5mm}
\scalebox{0.830}{\includegraphics{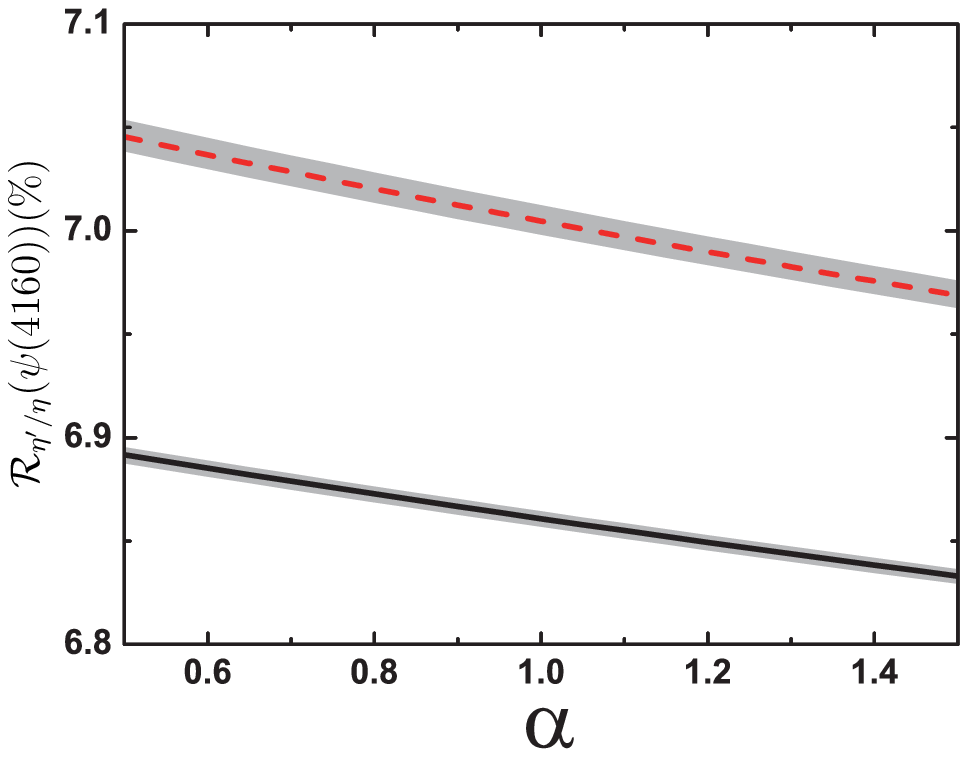}} %
\caption{{(color online) The $\alpha$ dependence of the coupling
constants $g_{\psi(4160)J/\psi\eta^\prime}^{D^{(*)}D^{(*)}}$ (upper panel) and the ratio of $\mathcal{B}(\psi(4160)\to
J/\psi \eta^\prime)$ to $\mathcal{B}(\psi(4160)\to J/\psi \eta)$
(lower panel). The solid and dashed curves with errors in the lower panel
are the results when the relative sign of coupling constants $g_{\psi(4160)D^\ast D^\ast}$ and $g_{\psi(4160) D^\ast D}$ is taken to be plus and minus,
respectively. \label{Fig:branching_2} }}
\end{figure}

\section{Numerical Results and Discussion }
The coupling constants $g_{\psi^\prime J/\psi \eta}^{D^{(*)}D^{(*)}}$
defined in Eqs.~(\ref{Eq:Lorentz0}-(\ref{Eq:Lorentz}) can be evaluated by estimating the
corresponding loop integrals.
In Fig.
\ref{Fig:coupling}, we show the absolute values of the coupling
constants corresponding to different intermediates.
{The grey bands are the uncertainties
of $g_{\psi^\prime J/\psi \eta}^{D^{(*)}D^{(*)}}$, which are
resulted from the errors of the coupling constants of
$\psi(4040)/\psi(4160)$ interacting with charmed meson pair.
We notice that there exits large uncertainty for
$g^{D^\ast D^\ast}_{\psi(4040) J/\psi \eta}$ compared with $g^{D^\ast D}_{\psi(4040) J/\psi \eta}$ and $g^{D D}_{\psi(4040) J/\psi \eta}$. For $\psi(4160)$, the coupling $g^{D
\bar{D}}_{\psi(4160) J/\psi \eta}$ has a large error as shown in Fig. \ref{Fig:coupling}.}
In the $\psi(4040)
\to J/\psi \eta$ process, the coupling constant corresponding to
$D^\ast \bar{D} +h.c$ channel is about 3 times larger than that for
$D\bar{D}$ channel, while the one for $D^\ast \bar{D}^\ast$ is
nearly one order larger than that for $D\bar{D}$ channel. The
coupling constants for $\psi(4160) J/\psi \eta$ behave very similar to
those for $\psi(4040) \to J/\psi \eta$. The dominant
contribution comes from the process of the $D^\ast \bar{D}^\ast$
channel while the one corresponding to $D \bar{D}$ is very small and
can be neglected.

{Having the couplings evaluated above, one can estimate the
branching ratios of $\psi(4040/4160) \to J/\psi \eta$. Since the
signs of the coupling constants for $\psi^\prime D^{(\ast)}
D^{(\ast)}$ $(\psi^\prime=\{\psi(4040), \psi(4160)\})$
interactions are unknown, we have to consider the sign effects on
the results, where the relative signs among $g_{\psi^\prime
\mathcal{D}^{(\ast)} {D}^{(\ast)}}$ directly result in those among
$g_{\psi^\prime J/\psi \eta}^{D^{(*)}D^{(*)}}$ listed in Eqs.
(\ref{Eq:Lorentz0})-(\ref{Eq:Lorentz}). As shown in Fig.
\ref{Fig:coupling}, the contributions via the intermediate
$D\bar{D}$ channel to $\psi(4040)/\psi(4160)\to J/\psi\eta$
processes can be ignored compared with those of other intermediate
channels, which enables us to further simplify our calculation.
Here, we take the same sign for $g_{\psi^\prime \mathcal{DD}}$ and
$g_{\psi^\prime \mathcal{D}^\ast \mathcal{D}}$. Next, we must
consider two typical cases, i.e, the relative sign between
$g_{\psi^\prime \mathcal{D}^\ast \mathcal{D}}$ and $g_{\psi^\prime
\mathcal{D}^\ast \mathcal{D}^\ast}$ is plus or minus. Finally, we
present the results of the branching ratios for
$\psi(4040)/\psi(4160)\to J/\psi\eta$ in Fig. \ref{Fig:branching}.
Because of the uncertainties of $g_{\psi^\prime
J/\psi\eta}^{D^{(*)}D^{(*)}}$, the results in Fig.
\ref{Fig:branching} are drawn using their central values with
errors.}

\iffalse The black solid and red dashed curves in Fig.
\ref{Fig:branching} are obtained by the central value of the
coupling constants in Fig. \ref{Fig:coupling}. The grey band in the
upper panel of Fig. \ref{Fig:branching} are the error of the
branching ratio caused by the uncertainties of $g_{\psi(4040) J/\psi
\eta}^{D^\ast D^\ast}$, which is the largest one among
$g_{\psi(4040) J/\psi \eta}^{D^{(\ast)} D^{(\ast)}}$ . The errors
for different sign of $g_{\psi^\prime D^{\ast} D^{\ast}}$ overlap to
each other and form a large band. While for $\psi(4160)$, the
coupling $g_{\psi(4160) J/\psi \eta}^{D D}$ has a largest relative
error, however, it is still nearly one order smaller than
$g_{\psi(4160) J/\psi \eta}^{D^\ast D^\ast}$, the uncertain of the
branching ratio caused by $g_{\psi(4160) J/\psi \eta}^{D D}$ is less
than $5\%$. We present the error caused by $g_{\psi(4160) J/\psi
\eta}^{D^\ast D^\ast}$ in the lower panel of Fig.
\ref{Fig:branching} since it dominantly contribute to the branching
fraction. \fi

{To compare our results with the
experimental measurements, we have presented the experimental value
of $\psi(4040)\to J/\psi\eta$ given by BESIII \cite{Ablikim:2012ht}
and Belle \cite{Bruce:2012,Wang:2012bg} in Fig. \ref{Fig:branching}.
Here, the branching ratios $\mathcal{B}(\psi(4040)\to J/\psi \eta)$
corresponding to solutions I and II from Belle are $(0.56 \pm 0.10
\pm 0.17) \%$ and $(1.30 \pm 0.15 \pm 0.24) \%$ \cite{Bruce:2012,
Wang:2012bg}, respectively, while the BESIII measurement gives
$\mathcal{B}(\psi(4040)\to J/\psi \eta )=(5.2 \pm 0.5 \pm 0.2 \pm
0.5)\times 10^{-3}$ \cite{Ablikim:2012ht}. Since the second solution
of the Belle Collaboration is not consistent with the one of the
BESIII and CLEO Collaboration, we consider only the first solution
from the Belle Collaboration and results of the BESIII
Collaboration, which are the cyan and yellow bands in Fig.
\ref{Fig:branching}. Even though there exists uncertainty in our
theoretical results for $\psi(4040)\to J/\psi\eta$ due to the large
error and undetermined sign of $g_{\psi(4040) D^\ast D^\ast}$, we
can find that our theoretical curves overlap with the experimental
measurements in a reasonable parameter region for $\alpha$. Thus, we
can conclude that the meson loop effects can provide sizable
contributions to $\psi(4040) \to J/\psi \eta$ and explain the
experimental data for this process.}

{In this work, we also study
$\psi(4160) \to J/\psi \eta$ decay. The results shown in Fig.
\ref{Fig:branching} indicate that we can well explain the Belle's
data, i.e., our results overlap with the experiment measurement in
the range of $0.50<\alpha<0.80$ and $0.53 <\alpha<1.20$ under two
typical cases of relative signs, where the values $\alpha$ obtained
are reasonable. In addition, we also notice that these $\alpha$
ranges for $\psi(4040)\to J/\psi\eta$ and $\psi(4160)\to J/\psi\eta$
overlap with each other, which can be due to the similarity existing
in these two processes. This observation also gives an extra test to
our theoretical calculation and the hadron loop effects on
$\psi(4040/4160)\to J/\psi\eta$. }

\iffalse

{The results shown in Fig.
\ref{Fig:branching}, we also notice that there exist overlap of the
obtained $\alpha$ values for explaining the large branching ratios
of $\psi(4040)\to J/\psi\eta$ and $\psi(4160)\to J/\psi\eta$ decays.
Considering the similarity between $\psi(4040/4160)\to J/\psi\eta$
and $\psi(4040/4160)\to J/\psi\eta^\prime$, we can apply these
obtained $\alpha$ range to estimate the branching ratio of
$\psi(4040)\to J/\psi\eta^\prime$ and $\psi(4160)\to
J/\psi\eta^\prime$, where $\alpha$ is constrained as $**\sim **$. By
this way, the uncertainty of our predictions is further controlled.
In the following, we present the calculated results for
$\psi(4040)\to J/\psi\eta^\prime$ and $\psi(4160)\to
J/\psi\eta^\prime$ decays.} \fi

{Other than the $\eta$ transition
between $\psi(4040/4160)$ and $J/\psi$, we also calculate the meson
loop contributions to $\psi(4160) \to J/\psi \eta^\prime$. This
prediction can be an important test to the meson loop mechanism
proposed in this work. The formalism for $\psi(4160) \to J/\psi
\eta^\prime$ is similar to that for $\psi(4160) \to J/\psi \eta$,
where we only need to make some replacements of the corresponding
parameters and masses. Similar to the way applied to Fig.
\ref{Fig:coupling}, we present the obtained coupling constants
$g_{\psi(4160)J/\psi\eta^\prime}^{D^{(*)}D^{(*)}}$ in the upper
panel of Fig. \ref{Fig:branching_2}, where the $\psi(4160) \to
J/\psi \eta^\prime$ process via intermediate $D^\ast \bar{D}^\ast$
state is dominant, while the $D\bar{D}$ channel can be negligible.
With these obtained coupling constants, we predict the branching
ratio for $\psi(4160) \to J/\psi \eta^\prime$, where we also
consider two typical cases since we consider the effects of
different signs of
$g_{\psi(4160)\mathcal{D}^{(*)}\mathcal{D}^{(*)}}$, which is similar
to the treatment when calculating $\psi(4040)\to J/\psi \eta$. In
Fig. \ref{Fig:branching_2}, we do not directly give the branching
ratio for $\psi(4160) \to J/\psi \eta^\prime$. Instead, we show the
ratio of the branching ratios for $\psi(4160) \to J/\psi
\eta^\prime$ and $\psi(4160) \to J/\psi \eta$, where this ratio is
denoted as $\mathcal{R}_{\eta^\prime/\eta}(\psi(4160))$. In Fig.
\ref{Fig:branching_2}, we show the variation of
$\mathcal{R}_{\eta^\prime/\eta}(\psi(4160))$ in $\alpha$. Even in a
large region $0.5<\alpha<1.5$, the obtained ratio
$\mathcal{R}_{\eta^\prime/\eta}(\psi(4160))$ is $(6.96 \sim 7.05)
\%$ and $(6.83\sim 6.89) \%$ for two typical cases as mentioned
above. We find that this ratio is not strongly dependent on
$\alpha$. Therefore, in this approach, we can well control the
uncertainty of our prediction. Using the experimental data of
$\psi(4160)\to J/\psi\eta$ ($\mathcal{B}(\psi(4160) \to J/\psi \eta)
= 0.48 \pm 0.10 \pm 0.17 \%$) \cite{Bruce:2012,Wang:2012bg} and this
obtained ratio, we can predict $\mathcal{B}(\psi(4160) \to J/\psi
\eta^\prime) \simeq (2.0 \sim 4.8) \times 10^{-4}$, which is
consistent with the upper limit ( $< 5 \times 10^{-3}$) for
$\mathcal{B}(\psi(4160) \to J/\psi \eta^\prime)$ given by the CLEO
measurement with $90\%$ confidence level \cite{Coan:2006rv}.}

\section{Summary}

The charmonium above the threshold of
a pair of charmed mesons dominantly decays into charmed mesons, which
can couple with charmonium and light meson by
exchanging a proper charmed meson, such a mechanism, i.e, the meson
loop effect, is essential to understand the decay behavior of higher
charmonia above the thresholds in obtaining the enhanced decay amplitudes.

Stimulated by the experimental measurements by the Belle and
BESIII Collaborations
\cite{Bruce:2012,Wang:2012bg,Ablikim:2012ht}, we introduce the
meson loop mechanism to study the $\eta$ transitions between
$\psi(4040/4160)$ and $J/\psi$ in an effective Lagrangian
approach. The theoretical estimates have shown overlaps with the
experimental measurements in a reasonable parameter range. More
over, we have predicted the branching ratio of $\psi(4160) \to
J/\psi \eta^\prime$ of the order of $10^{-4}$, which can be tested
by future experiments. {We should emphasize that if we include
finite-width effects \cite{Giacosa:2007bn} of $\psi(4040)$ and
$\psi(4160)$ in this paper, we might be able largely to improve
our results, and in addition, we may calculate a nonzero decay
ratio of $\psi(4040)\to J/\psi\eta^\prime$ even though this is
kinematically forbidden. However, inclusion of the finite-width
effects introduces another ambiguity that is out of our control,
and hence we have not included this type of effects in this paper.
}

\iffalse We notice that the sum of final state masses for
$\psi(4160)\to J/\psi\eta^\prime$ is close to the mass of
$\psi(4160)$. Thus, drives us further to take into account the
effects arising from rather sizable finite width of $\psi(4160)$.
Thus, in Appendix we will present the relevant calculation for the
decays of $\psi(4040)\to J/\psi\eta^{(\prime)}$ and $\psi(4160)\to
J/\psi\eta^{(\prime)}$ by introducing the sizable finite widths of
$\psi(4040)$ and $\psi(4160)$. The results show that the ratio of
$\Gamma_{\psi(4160) \to J/\psi \eta^\prime}$ to
$\Gamma_{\psi(4160) \to J/\psi \eta}$ is about $ (14 \sim 19) \%$,
which is about $2\sim 3$ times larger than the results without
including the finite width effects of of $\psi(4160)$. With this
ratio, we can give the branching fractions of $\psi(4160) \to
J/\psi \eta^\prime$ is $(4 \sim 12) \times 10^{-4}$. That of
$\psi(4040)\to J/\psi\eta^\prime$ is also estimated as $(3\sim
6)\times 10^{-4}$.}\fi

Before closing this section, we would like to discuss the possible
extension of our work. We notice possible radiative decays of
$\psi(4040)/\psi(4160)$ into $\gamma\eta$ or $\gamma\eta^\prime$,
which are similar to the $\psi(4040)/\psi(4160)\to
J/\psi\eta^{(\prime)}$ processes discussed in this work, where
$\psi(4040)/\psi(4160)\to \gamma\eta^{(\prime)}$ can also occur
through triangle charmed-meson loops.
%Thus, as a future outlook, studying the
%$\psi(4040)/\psi(4160)\to \gamma\eta^{(\prime)}$ decays will be
%potentially interesting.
So far experiments have not yet observed $\psi(4040)/\psi(4160)\to
\gamma\eta^{(\prime)}$. Thus, theoretical estimate of the
branching ratios for $\psi(4040)/\psi(4160)\to
\gamma\eta^{(\prime)}$ may provide important information of
further experimental search for these interesting radiative decay
channels. What is more important is that $\psi(4040)/\psi(4160)\to
\gamma\eta^{(\prime)}$ can be an important test to the hadronic
loop effects on the higher charmonium decays.

\section*{Acknowledgment}

This project is supported by the National Natural Science
Foundation of China under Grant Nos. 11222547, 11175073, 11005129,
11035006, the Ministry of Education of China (FANEDD under Grant
No. 200924, SRFDP under Grant No. 20120211110002, NCET, the
Fundamental Research Funds for the Central Universities), the Fok
Ying Tung Education Foundation (No. 131006), and the West Doctoral
Project of Chinese Academy of Sciences.

\end{document}